\documentclass[prc,amsfonts,reprint, superscriptaddress,showpacs,
notitlepage,
longbibliography
]{revtex4-1} 

\usepackage{hyperref}
\hypersetup{colorlinks=true, linkcolor=blue, citecolor=red, urlcolor=blue}
\usepackage{tikz}
\usetikzlibrary{arrows}
\usetikzlibrary{decorations.pathmorphing}
\usepackage{hyperref}
\usepackage{amssymb}
\usepackage{amsmath}
\usepackage{color}
\usepackage{xcolor}
\usepackage{graphicx}
\usepackage{soul}
\usepackage{mathtools}
\usepackage{braket}
\usepackage{float}
\usepackage[normalem]{ulem}

\newcommand{\be}{\begin{equation}\begin{aligned}\label}{}
\newcommand{\ee}{\end{aligned}\end{equation}}
{}
{}
{}

\newcommand{\comment}[1]{}
\DeclareMathSizes{10}{9}{6}{6}

\begin{document}

\title{Heat transport in the quantum  Rabi model: Universality and ultrastrong coupling effects}

\author{L. Magazz\`u}
\affiliation{Institute for Theoretical Physics, University of Regensburg, 93040 Regensburg, Germany}

\author{E. Paladino}
\affiliation{Dipartimento di Fisica e Astronomia Ettore Majorana, Universit\`a di Catania, Via S. Sofia 64, I-95123, Catania, Italy;\\
INFN, Sez. Catania, I-95123, Catania, Italy; and CNR-IMM, Via S. Sofia 64, I-95123, Catania, Italy}

\author{M. Grifoni}
\affiliation{Institute for Theoretical Physics, University of Regensburg, 93040 Regensburg, Germany}

\date{\today}

\begin{abstract}
Heat transport in the quantum Rabi model at weak interaction with the heat baths is controlled by the qubit-oscillator coupling. Universality of the linear conductance versus the temperature is found for $T\lesssim T_K$, with $T_K$ a coupling-dependent Kondo-like temperature. 
At low temperature, coherent heat transfer via virtual processes yields a $\sim T^3$ behavior with destructive interference in the presence of quasi-degeneracies in the spectrum. As the temperature increases, incoherent emission and absorption dominate and a maximum is reached at $T\sim T_K/2$. In the presence of a bias on the qubit, the conductance makes a transition from a resonant to a broad, zero-bias peak regime.
Parallels and differences are found compared to  the spin-boson model in [K. Saito and T. Kato, Phys. Rev. Lett. \textbf{111}, 214301 (2013)], where the qubit-bath coupling instead of the internal qubit-oscillator coupling rules thermal transport.  
\end{abstract}

\maketitle

Qubit-resonator systems, described by the celebrated quantum Rabi model~\cite{Rabi1936,Rabi1937}, constitute the building blocks of circuit-QED platforms~\cite{Blais2021}. They allow for tunable light-matter coupling strengths
ranging from the weak to the ultrastrong coupling (USC) and beyond, with the coupling of the same order or even exceeding the qubit/resonator frequencies~\cite{Forn-Diaz2010,Yoshihara2017,Forn-Diaz2018review,Kockum2019,Giannelli2024}. Recent experiments have demonstrated the control of photonic heat current between heat baths contacted via composite superconducting qubit-resonator junctions~\cite{Pekola2021, Gubaydullin2022, Upadhyay2023}. These can behave as heat valves~\cite{Ojanen2008, Ronzani2018}, and, when coupled asymmetrically, as heat diodes~\cite{Segal2005PRL, Segal2005, Segal2006, Senior2020, Bhandari2021}.
As such, platforms based on qubit-resonator systems appear to be ideal for demonstrating different heat transport regimes and effects with a single device. \\
\indent Inspired by the experiments, theory works have addressed quantum heat transport in the Rabi model and its generalizations.
Using a numerically exact technique, experimental results were reproduced  in~\cite{Xu2021}, and a benchmark for perturbative approaches in the system-bath coupling was provided. The presence of characteristic two-peak structures in the temperature dependence of the conductance is predicted at USC in~\cite{Yamamoto2021}. The impact of the qubit-resonator coupling mechanisms, defined by a mixing angle between longitudinal and transverse coupling, has been explored in~\cite{Chen2022}, showing that, varying the qubit-oscillator coupling, the maximum heat current is attained at different angles.\\
\indent The simpler setup of an individual qubit coupled directly to the baths embodies the spin-boson model (SBM)~\cite{Leggett1987,Weiss2012}. The thermal conductance $\kappa$ exhibits, in the Ohmic SBM, a scaling behavior as a function of $T/T_K$, with $T_K(\alpha)$ a characteristic Kondo temperature which depends on the coupling $\alpha$ to the Ohmic baths~\cite{Saito2013}.
At low temperatures, $T\ll T_K$, heat transport occurs via coherent processes yielding the universal power-law $\kappa \sim T^3$. At intermediate-to-high temperatures, incoherent absorption/emission processes  result in the coupling-dependent power law $\kappa \sim T^{2\alpha-1}$~\cite{Ruokola2011,Saito2013,Yang2014, Yamamoto2018}.\\
\indent In this work, we use a diagrammatic technique in Liouville space to show that the quantum Rabi model also exhibits distinct heat transport regimes and a scaling behavior with an appropriate Kondo temperature. 
In the case of \emph{weak} coupling to Ohmic reservoirs, $T_K$ only depends on the qubit-resonator coupling $g$. 
An intuition on why one would expect the obtained features in junctions weakly-coupled to the baths is provided by the exact mapping~\cite{Garg1985} of the dissipative Rabi model with Ohmic baths to a SBM with structured bath. Indeed, the qubit-resonator coupling enters the effective coupling of the qubit to the structured bath and impacts the dynamics~\cite{Goorden2004,Nesi2007NJP,Magazzu2021}. 
Following this picture, a zero-temperature localization transition at a critical value $g_c \sim 1/\sqrt{\alpha}$, has been predicted for the dissipative Rabi model~\cite{DeFilippis2023}. For the weak system-bath coupling considered in the present work, the transition occurs at extremely large values of $g$ which are not of relevance here.\\
\begin{figure}[ht!]
\begin{center}
\includegraphics[width=0.45\textwidth,angle=0]{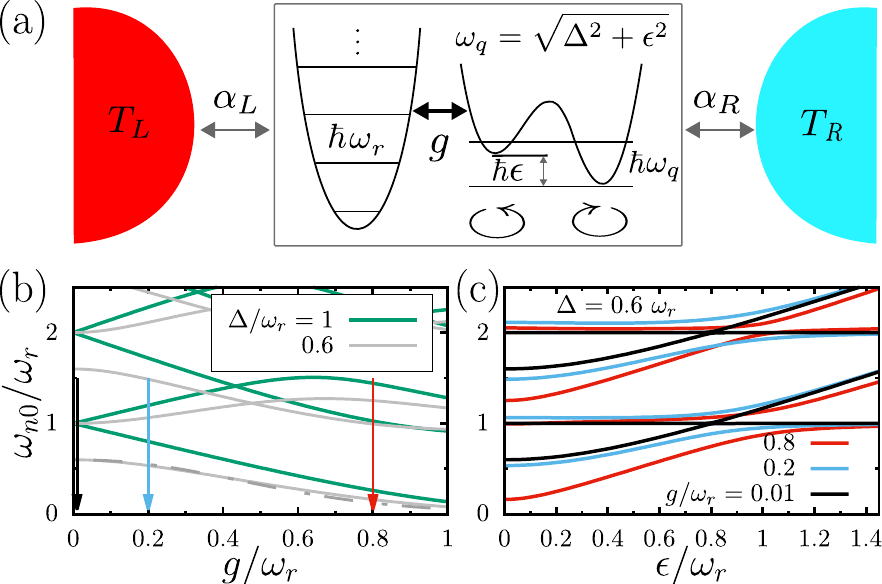}
\caption{\small{Heat transport setup and excitation spectrum of the Rabi model. (a) - A flux qubit-resonator system is weakly coupled to independent Ohmic baths. The circular arrows indicate left/right-circulating persistent current states. (b)-(c) - Excitation spectrum of the Rabi model \emph{vs.} qubit-resonator coupling strength at zero bias in (b) and \emph{vs.} the qubit bias at $\Delta=0.6~\omega_r$ in (c).}}
\label{fig_setup}
\end{center}
\end{figure}
\indent \emph{Setup.} - A scheme of the heat transport setup considered in this work is provided in Fig.~\ref{fig_setup}. The junction is formed of a  superconducting qubit coupled to a resonator. Exemplarily, we consider a flux qubit, where the qubit
frequency $\omega_q=\sqrt{\Delta^2+\epsilon^2}$ can be modulated by an applied bias $\epsilon$. The resonator is modeled as a quantum oscillator of frequency $\omega_r$. A simplified circuit scheme of the setup is provided in Appendix~\ref{implementation}. This system is described by the Rabi Hamiltonian which, in the basis of the qubit persistent current states $\{\ket{\circlearrowright },\ket{\circlearrowleft}\}$, has the form~\cite{Forn-Diaz2010,Yoshihara2017}
\begin{equation}
\hat H_{\rm Rabi}=-\frac{\hbar}{2}\left(\epsilon\hat \sigma_z + \Delta\hat \sigma_x \right) +\hbar\omega_r \hat a^\dagger \hat a + \hbar g\hat \sigma_z \left(\hat a^\dagger +\hat a \right)\;.
\label{H_Rabi}
\end{equation}
Here, $\hat a^\dag$ ($\hat a$) creates (annihilates) an excitation of the field mode in the resonator while \hbox{$\hat \sigma_z=\ket{\circlearrowright}\bra{\circlearrowright}-\ket{\circlearrowleft}\bra{\circlearrowleft}$} and \hbox{$\hat \sigma_x= \ket{\circlearrowright}\bra{\circlearrowleft}+\ket{\circlearrowleft}\bra{\circlearrowright}$} are the Pauli spin operators.  The angular frequency $g$ quantifies the coupling strength. The nonperturbative USC regime is attained for \hbox{$0.3\lesssim g/\omega_r\lesssim 1$}~\cite{Forn-Diaz2018review}.\\ 
\indent At weak-to-intermediate coupling, $g\lesssim 0.3$, the second-order Van Vleck perturbation theory (VVPT) in $g$~\cite{Hausinger2008} correctly describes the Rabi model for arbitrary $\Delta/\omega_r$, see Appendix~\ref{analytical}. Its weak coupling limit reproduces the rotating wave approximation (RWA), see Appendix~\ref{analytical}. On the other hand,
the generalized rotating wave approximation (GRWA)~\cite{Irish2007,Zhang2013} is valid for arbitrary $g$ and $\Delta/\omega_r\lesssim 1$. 
As we shall see, the frequency $\omega_{10}=(E_1-E_0)/\hbar$ gives the Kondo-like temperature via $T_K=\hbar\omega_{10}/k_B$.
Within the GRWA, the frequency gap $\omega_{10}$ at zero bias is approximated by
\begin{equation}
\begin{aligned}\label{eigensys_GRWA}
\omega_{10}&\simeq\;\Delta e^{-\tilde\alpha /2} - \frac{\delta}{2}
-\frac{1}{2}\sqrt{\delta^2+(\Delta e^{-\tilde\alpha /2}\tilde\alpha)^2}\;,
\end{aligned}
\end{equation}
where $\delta  :=\Delta e^{-\tilde\alpha /2}(2-\tilde\alpha)/2  -\omega_r$ and where $\tilde\alpha:=(2g/\omega_r)^2$ is the \emph{internal} effective coupling, see Appendix~\ref{analytical}.
The dashed-dotted line in Fig.~\ref{fig_setup}(b) reproduces $T_K$, according to Eq.~\eqref{eigensys_GRWA}, for $\Delta=0.6~\omega_r$.\\ 
\indent  We model the two-bath setup within the Caldeira-Leggett Hamiltonian~\cite{Caldeira1981, Caldeira1983, Magazzu2021}
\be{H_Caldeira_Leggett2baths}
\hat H=&\;\bar{H}_{\rm Rabi}+\sum_{l=R,L}( \hat H_{l} + \hat{Q}_l\hat{B}_l)\:,
\ee
where the second and third terms collect the baths' Hamiltonians $\hat H_{l}=\sum_{j}\hbar\omega_{lj} \hat b_{lj}^\dag \hat b_{lj}$ and the coupling of the system with the individual bath modes, respectively, 
via the bath displacement operators $\hat{B}_l\equiv \sum_j \hbar\lambda_{lj}(\hat b_{lj}+\hat b_{lj}^{\dag})$. The coupling is mediated by the dimensionless system operators $\hat{Q}_L=\hat a+\hat a^\dag$ and $\hat{Q}_R=\hat \sigma_z$. 
\begin{figure}[ht!]
\begin{center}
\includegraphics[width=0.47\textwidth,angle=0]{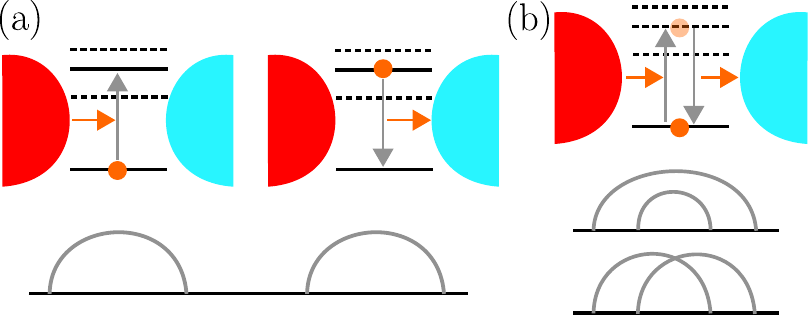}
\caption{\small{Relevant mechanisms for heat transfer, starting and ending in the ground state of the Rabi model, and their diagrams. Sequential (a) and coherent (b) transport and the corresponding diagrammatic contributions to the current kernels $\hat{\mathcal{K}}_{{\rm I}l}^{(2)}$ and $\hat{\mathcal{K}}_{{\rm I}l}^{(4)}$, respectively. The orange bullets indicate occupation of an energy level of the Rabi model. }}
\label{diagrams}
\end{center}
\end{figure}
The baths and their interaction with the system are collectively described by the spectral density functions $J_l(\omega)=\sum_{j=1}^{N}\lambda_{lj}^2\delta(\omega-\omega_{j})$. In the continuum limit, we assume the Ohmic-Drude form $J_l(\omega)=\alpha_l\omega/(1+\omega^2/\omega_c^2)$,  where $\omega_c$ is a high-frequency cutoff and $\alpha_l$ measures the coupling to bath $l$. Finally,  $\bar{H}_{\rm Rabi}=\hat H_{\rm Rabi}+\sum_l \hat{A}_l^2$ incorporates the bath-induced renormalization $\hat{A}_l^2=\hbar\int_{0}^{\infty}d\omega[J_l(\omega)/\omega]\hat{Q}_l^2$~\cite{Weiss2012}.\\
\indent \emph{Quantum heat transport.} -  The heat current $I_l(t)$ to bath $l$ is defined as the expectation value of the current operator $\hat{I}_l=d \hat H_{l}(t)/dt={\rm i} \hat{Q}_l\sum_{j}\lambda_{lj}\hbar\omega_{lj}[\hat b_{lj} - \hat b_{lj}^\dag]$, namely 
$I_l(t)=\big\langle \hat{I}_l(t) \big\rangle$.
For its evaluation, we have developed an exact diagrammatic approach in Liouville space~\cite{companion} which yields for the steady-state reduced density matrix and heat current to bath $l$ the equations
\be{current}
0=\mathcal{L}_{\rm Rabi}\rho^\infty+\tilde{\mathcal{K}}(0)\rho^\infty\quad\text{and}\quad I_l ={\rm Tr}[\tilde{\mathcal{K}}_{{\rm I}l}(0)\rho^\infty]\;.
\ee 
Here, $\mathcal{L}_{\rm Rabi}$ is the Liouvillian superoperator associated to the Rabi Hamiltonian in Eq.~\eqref{H_Rabi}, while $\tilde{\mathcal{K}}(0)$ and $\tilde{\mathcal{K}}_{{\rm I}l}(0)$ 
are the Laplace-transformed dynamical and current kernels, respectively.
The linear conductance follows as
$\kappa=\lim_{\Delta T \to 0}\partial_{\Delta T}I_l
$, where $\Delta T$ is the temperature difference between the baths.\\
\begin{figure}[ht!]
\begin{center}
\includegraphics[width=0.92\linewidth,angle=0]{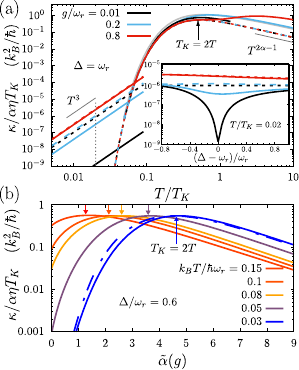}
\caption{\small{Scaling behavior and interference effects in the conductance. (a) - Linear conductance \emph{vs.} temperature for  three values of the coupling $g$ at $\epsilon=0$. 
The dashed lines reproduce $\kappa_{\rm TLS}$. Inset - conductance \emph{vs.} detuning at low temperature, see the dotted vertical line in the main panel. At resonance, for weak coupling, the first and second excited states are quasi-degenerate (see Fig~\ref{fig_setup}(b)). At low temperature, this entails a large suppression of $\kappa^{(4)}$ compared to $\kappa^{(4)}_{\rm TLS}$ due to multi-level interference effects. In the USC regime, where the lowest energy doublet is well-separated from higher levels, a scaling behavior is observed. There, $\kappa^{(4)}\simeq\kappa^{(4)}_{\rm TLS}$. At low temperature the sequential contribution is exponentially suppressed and $\kappa^{(2)}\simeq\kappa^{(2)}_{\rm TLS}$ for all values of $g$. 
(b) - Linear conductance \emph{vs.} the effective coupling $\tilde{\alpha}(g)=(2g/\omega_r)^2$ for different temperatures (solid lines) at $\epsilon=0$ and $\Delta=0.6~\omega_r$. Dashed-dotted line: Analytical result for $\kappa_{\rm TLS}^{(2)}$ within the GRWA. 
For both panels,
$\eta :=2\pi Q_{L,01}^2 Q_{R,01}^2/(Q_{L,01}^2+Q_{R,01}^2)$
and the positions of the maxima are found using Eq.~\eqref{kappa2Ohmic}.}}
\label{fig_conductance_vs_T_alpha}
\end{center}
\end{figure}
\indent At weak Ohmic coupling $\alpha$, heat transport is dominated by processes of second- and fourth-order in the system-bath coupling, depending on the temperature regime, see Fig.~\ref{diagrams}. The former describe sequential absorption and emission of bath excitations which dominate heat transport down to temperatures $T\lesssim T_K$. For $T\ll T_K$ though, due to the exponentially suppressed excited-state populations,  fourth-order processes constitute the dominant mechanism.  In this case, heat transfer is mediated by \emph{virtual} transitions to excited states, a mechanism dubbed ``cotunneling" in analogy with the electron transport counterpart~\cite{Ruokola2011,Yamamoto2018,Bhandari2021}.\\ 
\indent \emph{Universality, scaling, and interference.} - Numerical and analytical results for the linear conductance $\kappa$ are shown in Figs.~\ref{fig_conductance_vs_T_alpha} and~\ref{fig_conductance_vs_Delta_eps}. 
We consider identical Ohmic baths with $\alpha_L=\alpha_R\equiv\alpha=10^{-3}$, sufficiently small for a perturbative approach to be appropriate, and cutoff frequency $\omega_c=5~\omega_r$. Numerical results are obtained with truncation of the oscillator's Hilbert space to the first 10 levels.\\ 
\indent Figure~\ref{fig_conductance_vs_T_alpha}(a) shows the conductance scaled with $T_K$ as a function of $T/T_K$, for the three values of the qubit-resonator coupling highlighted in Fig~\ref{fig_setup}. 
The conductance exhibits a universal $T^3$ behavior at low temperature, followed by an exponential increase and a maximum at $T\simeq T_K/2$.  In the USC regime, the curves display a scaling behavior, namely they do not depend on $\omega_q/\omega_r$ for $T\lesssim T_K$~\cite{companion}. 
This behavior is similar to the one found for the SBM in~\cite{Saito2013} (see Fig.~2 there) when increasing the qubit-bath coupling. The similarity can be understood in terms of the clear separation in the excitation spectrum at intermediate to large values of $g$. At small $g$, the excited spectrum of the Rabi model exhibits a doublet structure, as seen in Fig.~\ref{fig_setup}(b). This degeneracy  results in destructive interference and in turn in the suppression of the low-temperature conductance by several orders of magnitude, as seen in the inset of Fig.~\ref{fig_conductance_vs_T_alpha}(a).
Analogously, in the intermediate-to-high temperature regime the
second-order conductance $\kappa^{(2)}$ in Fig.~\ref{fig_conductance_vs_T_alpha}(a) is suppressed in the presence of quasi degeneracies due to nonvanishing steady-state coherences,
see~\cite{companion} for details. This can be seen by comparing the black-solid curve (partial secular master equation~\footnote{due to a truncation to 5 levels performed here, the partial secular master equation is not accurate at high temperature and therefore not shown in this regime.}) with the light-gray curve (full secular approximation). Obviously, these features are not observed in the SBM.\\ 
\indent Returning to the USC regime, transport for $T\lesssim T_K$ is dominated by a two-level system (TLS) truncation of the full Rabi model whose frequency $\omega_{10}$ is negligibly renormalized by the weak coupling to Ohmic baths.
This is supported by the observation that, at USC, the Rabi model has a low-energy spectrum similar to the one of a fluxonium qubit with enhanced tunneling barrier~\cite{Manucharyan2017}. 
The second-order conductance of this effective TLS  reads
\be{kappa2Ohmic}
\kappa^{(2)}_{\rm TLS}
&=\alpha \eta k_B^2 T_K\frac{(T_K/T)^2}{2\hbar\sinh(T_K/T)}\;,
\ee
where $\eta :=2\pi Q_{L,01}^2 Q_{R,01}^2/(Q_{L,01}^2+Q_{R,01}^2)$ is a dimensionless asymmetry factor~\footnote{Notice that the matrix elements of the operators $\hat{Q}_l$ on the states $\ket{0},\ket{1}$ depend on the coupling $g$.}.
For temperatures $T < T_K$, Eq.~\eqref{kappa2Ohmic} gives $
\kappa^{(2)}_{\rm TLS}/T_K
\propto \alpha\eta
(T_K^2/T^2) e^{-T_K/T}$,
yielding for the position of the maxima the condition $T=T_K/2$. This also reproduces the position of the maxima in the full Rabi model, as shown in both panels of Fig.~\ref{fig_conductance_vs_T_alpha}.
At high temperatures, the TLS approximation breaks down at any $g$, since more levels of the Rabi model can contribute in the transport: This entails a departure from the TLS behavior $\sim T^{2\alpha-1}$~\cite{Saito2013,Yamamoto2018}.\\
\indent In the low-temperature regime, where essentially only the ground state of the Rabi system is populated, we find for the fourth-order conductance the  form~\cite{companion} 
\be{}
\kappa^{(4)}\simeq &
\frac{32\alpha^2\pi^5 k_B^4 T^3}{15\hbar^3}\sum_{k,m(\neq 0)}
\frac{Q_{R,m0}Q_{L,0m}Q_{L,0k}Q_{R,k0}}{\omega_{m0}\omega_{k0}}\;.
\ee
The above expression gives a $T^3$ behavior and accounts for interference effects in the presence of quasi degeneracies. Indeed at resonance and weak coupling, $Q_{L,01}=Q_{L,02}$ while $Q_{R,01}=-Q_{R,02}$, see Appendix~\ref{analytical}.
In the TLS truncation, the known expression is recovered $\kappa_{\rm TLS}^{(4)}/T_K \propto  Q_{L,10}^2Q_{R,10}^2(T/T_K)^3$, which does not display the multi-level interference effects. \\
\indent In Fig.~\ref{fig_conductance_vs_T_alpha}(b)  the conductance is depicted \emph{vs}. the effective coupling strength $\tilde\alpha$ for different temperatures (in a range where the conductance is dominated by $\kappa^{(2)}$). Numerical evaluations in the full Rabi model are compared to $\kappa_{\rm TLS}^{(2)}$ obtained within the GRWA. Remarkably, the results are qualitatively similar to those obtained in~\cite{Saito2013} for the conductance of a qubit \emph{vs.} the qubit-bath coupling strength. 
For large $\tilde\alpha$, the effective frequency $\omega_{10}$ is suppressed, so that the high-temperature expansion 
$\kappa^{(2)}_{\rm TLS}\propto \alpha\eta T_K^2/T$ is appropriate. 
\begin{figure}[ht!]
\begin{center}
\includegraphics[width=0.98\linewidth,angle=0]{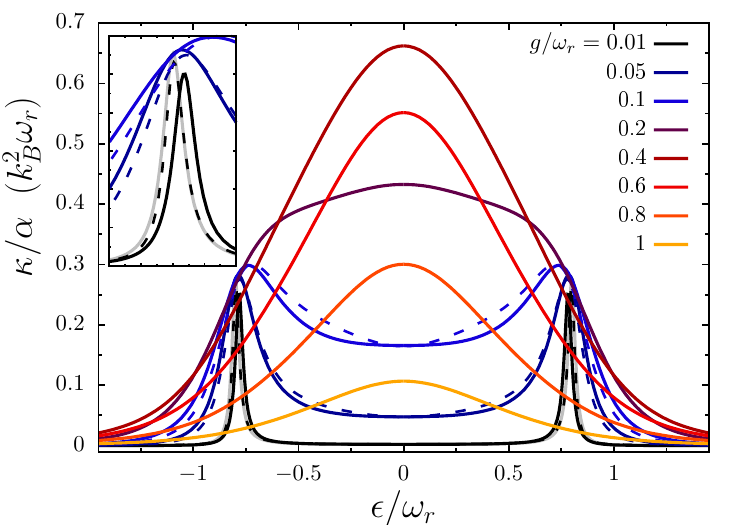}
\caption{\small{Transition from resonant to USC regime upon increasing the internal coupling $g$. At weak coupling, the conductance \emph{vs} qubit bias exhibits two sharp peaks at  resonance, $\omega_q=\omega_r$. As the coupling is increased, a zero-bias peak emerges. The gray solid line shows the conductance for $g/\omega_r=0.01$ in the full secular approximation. The quasi-degeneracy at resonance for small $g$, see Fig~\ref{fig_setup}(c), gives, in the partial secular master equation (black solid line) a suppressed peak with a Lamb shift, as highlighted in the inset. Dashed lines: $\kappa^{(2)}$ in the full secular approximation within VVPT in $g$ and truncation to the first four levels. The temperature is $k_BT=0.2~\hbar\omega_r$ and  $\Delta/\omega_r=0.6$.}}
\label{fig_conductance_vs_Delta_eps}
\end{center}
\end{figure}
The GRWA expression for $\omega_{10}$ at zero bias gives $\omega_{10}\simeq \Delta\exp(-\tilde\alpha/2)$, and we obtain
$\kappa^{(2)}_{\rm TLS}
\propto 
\alpha \eta 
\Delta^2 e^{-\tilde\alpha}/T$,
which accounts for the exponential decay at large $\tilde\alpha$.\\
\indent In Fig.~\ref{fig_conductance_vs_Delta_eps}, we provide numerical and semi-analytical (VVPT in $g$) results for the thermal conductance \emph{vs.} the detuning between qubit and resonator. Specifically, the effect of a flux bias $\epsilon$ applied to the qubit is considered for $\Delta=0.6~\omega_r$.
A transition from a resonant behavior at weak coupling to a broadened structure in the USC regime is observed. 
The curves show suppressed conductance at weak coupling, except for the peaks at resonance, $\omega_q=\omega_r$. 
Quasi-degeneracies in the spectrum at resonance produce non-vanishing coherences at the steady state. As a result, the conductance peaks at weak coupling, as calculated within the partial secular approximation, show a suppression and a Lamb shift with respect to the one in the full secular approximation~\cite{Cattaneo2019,Ivander2022,companion}.
Upon increasing $g$, the conductance peaks broaden and move towards lower frequencies, \hbox{$\omega_q <\omega_r$}. Entering the nonperturbative USC regime of qubit-resonator coupling, a single, zero-bias peak emerges.\\
\indent  The behavior of the conductance in Fig.~\ref{fig_conductance_vs_Delta_eps} can be accounted for by considering the matrix elements of the coupling operators $Q_{l,01}$.
In the weak coupling regime, the RWA predicts for the effective frequency $\omega_{10}=\omega_q-\delta+\Omega/2$, 
with $\Omega:=\delta - \sqrt{\delta ^2 + 4g_x^2}$, detuning $\delta:=\omega_q-\omega_r$, and $ g_x:= g\Delta/\omega_q$. The relevant matrix elements are approximated by $Q_{L,01}={\rm v}$ and $ Q_{R,01}={\rm u} \Delta/\omega_q$, where ${\rm u} :=\;\Omega/\sqrt{\Omega^2+4g_x^2}$, and ${\rm v} :=\;-2g_x/\sqrt{\Omega^2+4g_x^2}$.
Off-resonance, $\delta\gg g$, the conductance is suppressed as ${\rm v} \sim 0$, whereas at resonance \hbox{${\rm v}={\rm u}=-1\sqrt{2}$}, giving a peak in the conductance. This result is independent of $g$, which accounts for the pinning of the conductance at the resonance condition for different (small) values of $\tilde\alpha(g)$.
Increasing the coupling, the peaks broaden and shift towards lower frequencies. This behavior is captured by VVPT in $g$, where the resonance condition is modified as $\omega_q=\omega_r-2g_x^2/(\omega_q+\omega_r)$, and is therefore attained at smaller $\omega_q$, the larger the coupling $g$. The GRWA is well-suited for the USC regime and accounts for the zero-bias peak. Indeed, it gives for the matrix element of the qubit coupling operator $Q_{R,01}\simeq \Delta e^{-\tilde\alpha/2}/\sqrt{\Delta^2 e^{-\tilde\alpha}+\epsilon^2}$, which is suppressed at finite bias, see Appendix~\ref{analytical}.\\
\indent \emph{Conclusions.} - 
We have shown how a Rabi system weakly coupled to heat baths displays a rich variety of remarkable heat transport features.
They encompass a universal power-law behavior of the conductance, destructive interference at small qubit-oscillator coupling, and scaling together with zero-bias resonances in the USC regime.  The scaling behavior emerges when quantities are scaled with a Kondo-like temperature, which depends on the  qubit-resonator coupling. 
In order to account for both sequential and coherent heat transfer mechanisms, the thermal conductance has been calculated up to next-to-leading order using a diagrammatic technique in Liouville space. To this novel approach we dedicated the companion work~\cite{companion}. 
We expect that our predictions can be observed at experimentally  attainable regimes of weak system-bath and ultrastrong qubit-resonator coupling.

\section{Acknowledgements}
The authors thank A. Donarini, G. Falci, and J. Pekola for fruitful discussions.
LM and MG acknowledge financial support from BMBF (German Ministry for Education and Research), Project No. 13N15208, QuantERA SiUCs. The research is part of the Munich Quantum Valley, which is supported by the Bavarian state government with funds from the Hightech Agenda Bavaria. EP acknoweldges financial support from PNRR MUR project
PE0000023-NQSTI and from COST ACTION SUPQERQUMAP, CA21144.

\appendix

\begin{widetext}

\section{Superconducting circuit implementation}
\label{implementation}

The biased quantum Rabi model considered in this work, and shown in Fig. 1 of the main text, can be implemented in a superconducting quantum circuit.  A simplified scheme of the superconducting circuit implementation is presented in Fig.~\ref{circuit}. There, a flux qubit, biased by an applied magnetic flux $\phi_{\rm ext}$, is coupled to a superconducting LC oscillator via a shared, tunable inductance $L_c$~\cite{Forn-Diaz2010,Yoshihara2017,Magazzu2021}.

\begin{figure}[ht!]
\begin{center}
\includegraphics[width=0.6\textwidth,angle=0]{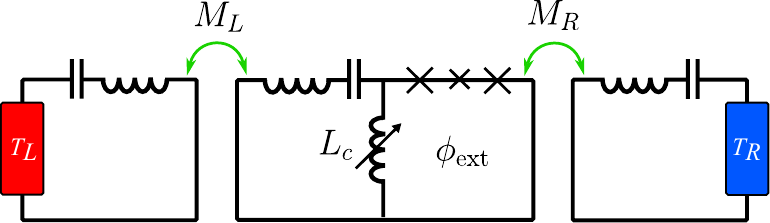}
\caption{\small{Superconducting circuit implementation of the heat transport setup where a qubit-oscillator system (described by the Rabi model) is the junction between the left and right bath.}}
\label{circuit}
\end{center}
\end{figure}
The  qubit-resonator coupling strength is estimated via $g/\omega_r = L_cI_p[2\omega_r (L_c + L_r)/\hbar]^{-1/2}$~\cite{Magazzu2021}. A coupling in the range $g/\omega_r=0.3-0.5$ can be attained with qubit persistent current $I_p=100$~nA, shared coupling inductance $L_c = 0.4$~nH, resonator inductance $L_r=3.5$~nH, and resonator frequency $\omega_r/2\pi=$  1~GHz. The resulting circuit is inductively coupled to the left and right heat reservoirs via the mutual inductances $M_L$ and $M_R$, respectively~\cite{Pekola2016, Karimi2017,Rasmussen2021}.
The heat reservoirs are realised by dissipative LC circuits ssumed to be out-of-resonance with the relevant frequencies of the qubit-oscillator system, so that effective Ohmic baths are seen by the latter. The inductive coupling between the resonator and the left bath realizes a coupling of the type $(a^\dag+a)(b_j^\dag + b_j)$ while the qubit-right bath inductive coupling yields an Hamiltonian term of the type $\sigma_x(b_j^\dag + b_j)$ (in the qubit energy basis)~\cite{Rasmussen2021}.

\section{Approximate diagonalization of the Rabi model}
\label{analytical}

In this section, we provide the analytical approximate solutions of the quantum Rabi model used in the main text. First, we detail the second-order Van Vleck perturbation theory. Its first-order truncation gives the rotating-wave approximation. Secondly, the generalized rotating-wave approximation is described, where the perturbation parameter is the  qubit splitting $\Delta$ dressed by the resonator. 

\subsection{Second-order Van Vleck perturbation theory (VVPT) in $g$}
\label{sec_VVPTg}

Let us define the frequency $\bar\omega=\omega_q+\omega_r$ and the couplings $g_z= g \epsilon/\omega_q$ and $g_x= g \Delta/\omega_q$. The latter are the longitudinal and transverse coupling, respectively, of the Rabi Hamiltonian expressed in the qubit energy basis.
Within second-order VVPT in the qubit-resonator coupling $g$~\cite{Hausinger2008} the eigenenergies $E_n=\hbar\omega_n$ of the Rabi model read
 ($n\geq 1$) 
\begin{equation}
\begin{aligned}\label{eigenenergies_VVPTg}
\omega_0 &=\,-\omega_q^{(1)}/2 - g_z^2/\omega_r\;,\\
\omega_{2n-1}&=\,-\frac{\omega_q^{(n)}}{2}+\frac{\delta_n}{2} +n\omega_r - \frac{g_z^2}{\omega_r}  - \frac{g_x^2}{\bar\omega}-\frac{1}{2}\sqrt{\delta_n^2+ 4n g_x^2 }\;, \\
\omega_{2n}&=\,-\frac{\omega_q^{(n)}}{2}+\frac{\delta_n}{2} +n\omega_r - \frac{g_z^2}{\omega_r}  - \frac{g_x^2}{\bar\omega}+\frac{1}{2}\sqrt{\delta_n^2+ 4n g_x^2 }\;,
\end{aligned}
\end{equation}
where $\omega_q^{(n)}=\omega_q+2ng_x^2/\bar\omega$ and $\delta_n=\omega_q^{(n)}-\omega_r$. 
Note that this indexing provides the correct ordering of the eigenfrequencies only for small enough $g$, $\epsilon$.
The corresponding eigenvectors are 
\begin{equation}
\begin{aligned}
\label{eigenstates_VVPTg}
|0\rangle &= |\widetilde{{\rm g},0}\rangle^{(2)}\,,\\
|2n-1\rangle &= {\rm u}_{n}^-|\widetilde{{\rm e},n-1}\rangle^{(2)} + {\rm v}_{n}^-|\widetilde{{\rm g},n}\rangle^{(2)}\;,\\
|2n\rangle &= {\rm u}_{n}^+|\widetilde{{\rm e},n-1}\rangle^{(2)} + {\rm v}_{n}^+|\widetilde{{\rm g},n}\rangle^{(2)}\;,
\end{aligned}    
\end{equation}
where the coefficients are given by
\begin{equation}
\begin{aligned}
\label{}
{\rm u}_{n}^\pm &:=\frac{\delta_n \pm \sqrt{\delta_n ^2+4n g_x^2}}{\sqrt{\left(\delta_n \pm \sqrt{\delta_n ^2+4n g_x^2}\right)^2+4n g_x^2}}\;,\qquad{\rm and}\qquad
{\rm v}_{n}^\pm  :=\frac{-2\sqrt{n}g_x}{\sqrt{\left(\delta_n \pm \sqrt{\delta_n ^2+4n g_x^2}\right)^2+4n g_x^2}}\;.
\end{aligned}    
\end{equation}
Note that in the main text we use the simplified notation ${\rm u}_{1}^-\equiv{\rm u}$ and ${\rm v}_{1}^-\equiv {\rm v}$.\\

The explicit form of the transformed states $ |\widetilde{e/g,n}\rangle^{(2)}$ in terms of the uncoupled energy eigenbasis $\{\Ket{{\rm g},n},\Ket{{\rm e},n}\}$ is
\be{}
\frac{|\widetilde{{\rm g},0}\rangle^{(2)}}{\mathcal{N}_{g,0}}&=\,
\Ket{{\rm g},0}+f(1)\Ket{{\rm e},0}+\frac{g_z}{\omega_r}\Ket{{\rm g},1}+\frac{g_x}{\bar\omega}\Ket{{\rm e},1}\;,\\
\frac{|\widetilde{{\rm g},n}\rangle^{(2)}}{\mathcal{N}_{g,n}}&=\,
-\sqrt{n}\frac{g_z}{\omega_r}\Ket{{\rm g},n-1}+\Ket{{\rm g},n}+f(n+1)\Ket{{\rm e},n}+\sqrt{n+1}\frac{g_z}{\omega_r}\Ket{{\rm g},n+1}+\sqrt{n+1}\frac{g_x}{\bar\omega}\Ket{{\rm e},n+1}\quad(n\geq 1)\;,\\
\frac{|\widetilde{{\rm e},0}\rangle^{(2)}}{\mathcal{N}_{e,0}}&=\,-f(1)\Ket{{\rm g},0}+\Ket{{\rm e},0}-\frac{g_z}{\omega_r}\Ket{{\rm e},1},\\
\frac{|\widetilde{{\rm e},n}\rangle^{(2)}}{\mathcal{N}_{e,n}}&=\,-\sqrt{n}\frac{g_x}{\bar\omega}\Ket{{\rm g},n-1}+\sqrt{n}\frac{g_z}{\omega_r}\Ket{{\rm e},n-1}-f(n+1)\Ket{{\rm g},n}+\Ket{{\rm e},n}-\sqrt{n+1}\frac{g_z}{\omega_r}\Ket{{\rm e},n+1}
\quad(n\geq 1)\;,\\
\ee
where $\mathcal{N}_i$ are the normalization factors and where $f(n):=(g_zg_x/\omega_r)[n/\omega_r-(n-1)/\bar\omega]$. 

\subsubsection{Zero bias, $\epsilon=0$}

At zero bias, the ground and first excited state read
\begin{equation}
\begin{aligned}
\label{eigenstates_VVPTg}
|0\rangle &\propto \,
\Ket{{\rm g},0}+\frac{g}{\Delta+\omega_r}\Ket{{\rm e},1}\;,\\
|1\rangle &\propto {\rm u}_{1}^-\Ket{{\rm e},0}+ {\rm v}_{1}^-   \left(
\Ket{{\rm g},1}+\frac{\sqrt{2}g}{\Delta+\omega_r}\Ket{{\rm e},2}\right)\;.
\end{aligned}    
\end{equation}
The matrix elements of the system's coupling operators ($\hat{Q}_L=\hat{a}+\hat{a}^\dag$ and $\hat{Q}_R=\sigma_z\epsilon/\omega_q-\sigma_x \Delta/\omega_q$, in the qubit energy basis) between the ground and first excited state read
\be{}
Q_{L,01}\propto{\rm v}_1^-\left(1+\frac{\sqrt{2}g}{\Delta+\omega_r}
\right)+{\rm u}_1^-
\frac{g}{\Delta+\omega_r}
\;,\qquad Q_{R,01}\propto{\rm u}_1^-
+
{\rm v}_1^-
\frac{g}{\Delta+\omega_r}\;.
\ee
The explicit expression for the coefficients is
\begin{equation}
\begin{aligned}
\label{}
{\rm u}_1^- &:=\frac{\delta_1 - \sqrt{\delta_1 ^2+4 g^2}}{\sqrt{\left(\delta_1 - \sqrt{\delta_1 ^2+4 g^2}\right)^2+4 g^2}}\;,\qquad{\rm and}\qquad
{\rm v}_{1}^-  :=\frac{-2g}{\sqrt{\left(\delta_1 - \sqrt{\delta_1 ^2+4 g^2}\right)^2+4 g^2}}\;,
\end{aligned}    
\end{equation}
where $\delta_1=\Delta+2g^2/(\Delta+\omega_r)-\omega_r$. Note that, at zero bias, these coefficients peak at the value of $\Delta<\omega_r$ for which the condition $\delta_1=0$ is satisfied.

\subsection{Rotating-wave approximation (RWA)}
\label{sec_RWA}

To first order in the coupling $g$, the  Van Vleck perturbation theory expressions for the eigensystem of the quantum Rabi model, Eqs.~\eqref{eigenenergies_VVPTg}-\eqref{eigenstates_VVPTg}, reproduce the RWA results 
\begin{equation}
\begin{aligned}
\label{JCMspectrum}
\omega_0=&-\omega_q/2\;,\\
\omega_{2n-1}=& \left(n-\frac{1}{2} \right)\omega_r - \frac{1}{2}\sqrt{\delta^2 + n 4 g_x^2}\;, \\
\omega_{2n}=& \left(n-\frac{1}{2} \right)\omega_r + \frac{1}{2}\sqrt{\delta^2 + n 4g_x^2}
\end{aligned}
\end{equation}
($n\geq 1$), with the detuning $\delta$ defined as $\delta:=\omega_q-\omega_r$ and $g_x= g\Delta/\omega_q$. The corresponding eigenstates, in the energy basis of the uncoupled system, are
\begin{equation}
\begin{aligned}
\label{JCMeigenstates}
|0\rangle &= |{\rm g},0\rangle\;,\\
|2n-1\rangle &= {\rm u}_{n}^-|{\rm e},n-1\rangle + {\rm v}_{n}^-|{\rm g},n\rangle\;,\\
|2n\rangle &= {\rm u}_{n}^+|{\rm e},n-1\rangle + {\rm v}_{n}^+|{\rm g},n\rangle\;,
\end{aligned}    
\end{equation}
with coefficients
\begin{equation}
\begin{aligned}\label{}
{\rm u}_n^\pm &:=\;\frac{\delta \pm \sqrt{\delta ^2 + n4g_x^2}}{\sqrt{\left(\delta \pm \sqrt{\delta^2+n4g_x^2}\right)^2+n4g_x^2}}\;,\qquad{\rm and}\qquad
{\rm v}_n^\pm :=\;\frac{-\sqrt{n}2g_x}{\sqrt{\left(\delta \pm \sqrt{\delta^2+n4g_x^2}\right)^2+n4g_x^2}}\,.
\end{aligned}
\end{equation}

Within the RWA, the relevant matrix elements of the system's coupling operators ($\hat{Q}_L=\hat{a}+\hat{a}^\dag$ and $\hat{Q}_R=\sigma_z\epsilon/\omega_q-\sigma_x \Delta/\omega_q$, in the qubit energy basis) read
\be{}
Q_{L,01}&={\rm v}_1^-\;,\qquad Q_{R,01}={\rm u}_1^-\frac{\Delta}{\omega_q}\\
Q_{L,02}&={\rm v}_1^+\;,\qquad Q_{R,02}={\rm u}_1^+\frac{\Delta}{\omega_q}\\
\ee
At resonance, $Q_{L,01}=Q_{L,02}=-1/\sqrt{2}$ while $Q_{R,01}=-Q_{R,02}=-1/\sqrt{2}$. If $\omega_{10}\simeq\omega_{02}$, the change of sign of the qubit matrix element yields the suppressed fourth-order conductance at weak coupling, see the main text.
Note that the RWA is the exact solution of the Jaynes-Cummings model~\cite{Shore1993}.

\subsection{Generalized rotating-wave approximation (GRWA)}
\label{appendix_GRWA}

Within the GRWA~\cite{Irish2007,Zhang2013}, the spectrum 
 $E_n=\hbar\omega_n$ of the biased Rabi model is approximated  by 
\begin{equation}
\begin{aligned}\label{eigensys_GRWA_app}
\omega_0 &=\;-\omega_{q,0}/2- g^2/\omega_r\;,\\
\omega_{2n-1}
&=\;-\frac{\omega_{q,n}}{2}+ \frac{\delta_n}{2}+n\omega_r - \frac{g^2}{\omega_r}
-\frac{1}{2}\sqrt{\delta_n^2+\Omega_{n}^2}\;,\\
\omega_{2n}
&=\;-\frac{\omega_{q,n}}{2}+ \frac{\delta_n}{2}+n\omega_r - \frac{g^2}{\omega_r}
+\frac{1}{2}\sqrt{\delta_n^2+\Omega_{n}^2}\;.
\end{aligned}
\end{equation}
Here, $\delta_n  :=(\omega_{q,n}+\omega_{q,n-1})/2  -\omega_r$, 
$\omega_{q,n}:=\;\sqrt{\tilde\Delta_{nn}^2 + \epsilon^2}$, and $\Omega_{n} :=\;\tilde\Delta_{nn-1}(c^+_n c^+_{n-1} +  c^-_n c^-_{n-1})$, where \hbox{$c^\pm_n :=\;\sqrt{(\omega_{q,n}\pm\epsilon)/2\omega_{q,n}}$}.
Here, we have introduced the dressed qubit gap $\tilde{\Delta}_{ij}= \Delta e^{-\tilde\alpha /2}\tilde\alpha^{(i-j)/2}\sqrt{j!/i!}\;\mathsf{L}^{i-j}_j(\tilde\alpha)$ ($i\geq j$), where $\tilde\alpha:=(2g/\omega_r)^2$ and where $\mathsf{L}^{k}_n$ are the generalized Laguerre polynomials defined by the recurrence relation
\be{}
\mathsf{L}^k_{j+1}(\tilde{\alpha})=\frac{(2j+1+k-\tilde{\alpha})\mathsf{L}^k_j(\tilde{\alpha})-(j+k)\mathsf{L}^k_{j-1}(\tilde{\alpha})}{j+1}\;,
\ee
with $\mathsf{L}^k_0(\tilde{\alpha})=1$ and $\mathsf{L}^k_1(\tilde{\alpha})=1+k-\tilde{\alpha}$. 
The corresponding energy eigenstates are
\be{eigenstates_GRWA_app}
\Ket{0}&=\ket{{\Psi_{+,0}}}\;,\\
\Ket{2n-1}&={\rm u}_n^{-}\ket{{\Psi_{-,n-1}}}+{\rm v}_n^{-}\ket{{\Psi_{+,n}}}\;,\\
\Ket{2n}&={\rm u}_n^{+}\ket{{\Psi_{-,n-1}}}+{\rm v}_n^{+}\ket{{\Psi_{+,n}}}\;,
\ee
with the weights given by
\begin{equation}
\begin{aligned}\label{GRWA_quantities}
{\rm u}_n^\pm &:=\;\frac{\delta_n \pm \sqrt{\delta_n ^2+\Omega_{n}^2}}{\sqrt{\left(\delta_n \pm \sqrt{\delta_n ^2+\Omega_{n}^2}\right)^2+\Omega_{n}^2}}\;,\qquad{\rm and}\qquad
{\rm v}_n^\pm :=\;\frac{-\Omega_{n}}{\sqrt{\left(\delta_n \pm \sqrt{\delta_n ^2+\Omega_{n}^2}\right)^2+\Omega_{n}^2}}\,.
\end{aligned}
\end{equation}
The states 
\be{Psi_pm}
\Ket{\Psi_{\pm,j}}
=\,c^\mp_j\Ket{-_z j_-}\pm c^\pm_j\Ket{+_z j_+}
\ee
are superpositions of the displaced states $\ket{\pm_z j_\pm}=\exp[ g\sigma_z(a-a^\dag)/\omega_r]\ket{\pm_z}|j\rangle\equiv \ket{\pm_z}D(\pm g/\omega_r)|j\rangle$, where $\{\ket{\pm_z}\}$ is the qubit localized basis, i.e. $\{\ket{+_z},\ket{-_z}\}\equiv\{\ket{\circlearrowright },\ket{\circlearrowleft}\}$, see the main text, and $D(x)=\exp[x(a-a^\dag)]$ is the displacement operator. 

The two-level system truncation of the Rabi model
 gives for the gap $\omega_{10} =\omega_1 - \omega_0$ of the effective TLS
\begin{equation}
\begin{aligned}\label{}
\omega_{10} &=\;\omega_{q,0} - \frac{\delta_1}{2}
-\frac{1}{2}\sqrt{\delta_1^2+\Omega_1^2}\;.
\end{aligned}
\end{equation}
The relevant matrix elements are $Q_{l,01}=\bra{0}\hat{Q}_l\ket{1}$, where $\hat{Q}_L=a+a^\dag$ and $\hat{Q}_R=\sigma_z$ (in the qubit localized basis). Using $D(x)aD^\dag(x)=a+x$ and $D(x)a^\dag D^\dag(x)=(D(x)a D^\dag(x))^\dag= a^\dag + x^*$,  
Eqs.~\eqref{eigenstates_GRWA_app}  and~\eqref{Psi_pm}
give
\be{matrix_elements_GRWA_app}
Q_{L,01} &=\, \frac{4 g}{\omega_r}{\rm u}_1^-  c^-_0 c^+_0 + {\rm v}_1^-(c^-_0 c^-_1 + c^+_0 c^+_1)
\;,\qquad
Q_{R,01} &=\,-2{\rm u}_1^- c^-_0 c^+_0
\;.
\ee

\subsubsection{Zero bias, $\epsilon=0$}

The first two energy levels and the corresponding eigenstates read
\begin{equation}
\begin{aligned}\label{}
\omega_0 &=\;-\frac{1}{2}\tilde\Delta_{00}-\frac{g^2}{\omega_r}\;,\,\quad\qquad\quad\qquad\qquad\qquad\qquad\qquad\qquad\qquad\qquad\qquad\qquad\; \Ket{0}=\ket{{\Psi_{+,0}}}\;,\\
\omega_1 &=\; \frac{1}{2}\left(\frac{\tilde\Delta_{00}-\tilde\Delta_{11}}{2} + \omega_r\right)-\frac{g^2}{\omega_r}-\frac{1}{2}\sqrt{\left(\frac{\tilde\Delta_{00}+\tilde\Delta_{11}}{2} - \omega_r\right)^2+(\tilde\Delta_{10})^2}\;,\qquad \Ket{1}={\rm u}_1^{-}\ket{{\Psi_{-,0}}}+{\rm v}_1^{-}\ket{{\Psi_{+,1}}}\;, 
\end{aligned}
\end{equation}
and, since $c^\pm_j(\epsilon=0)=1/\sqrt{2}$,
\be{}
\Ket{\Psi_{\pm,j}}
=\,\frac{1}{\sqrt{2}}\left(\Ket{-_z j_-}\pm \Ket{+_z j_+}\right)\,.
\ee
The gap $\omega_{10} =\omega_1 - \omega_0$ of the effective qubit is
\begin{equation}
\begin{aligned}\label{DeltaeffGRWAzerobias}
\omega_{10}
&=\;\tilde\Delta-\frac{1}{2}\left(\tilde\Delta-\omega_r -\frac{\tilde\alpha\tilde\Delta}{2}\right)
-\frac{1}{2}\sqrt{\left(\tilde\Delta-\omega_r -\frac{\tilde\alpha\tilde\Delta}{2}\right)^2+\tilde\alpha\tilde\Delta^2}
\;,
\end{aligned}
\end{equation}
where we have defined $\tilde\Delta:=\Delta\exp(-\tilde\alpha/2)=\tilde\Delta_{00}$ and used $\tilde\Delta_{11}=\Delta\exp(-\tilde\alpha/2)(1-\tilde\alpha)$ and $\tilde\Delta_{01}=\sqrt{\tilde\alpha}\tilde\Delta$. 
The matrix elements of $\hat{Q}_l$ in the basis $\{\Ket{0},\Ket{1}\}$ read
\be{QLR_GRWA}
Q_{L,01} &=\, \frac{2 g}{\omega_r}{\rm u}_1^-
+{\rm v}_1^-\;,\qquad\,
Q_{R,01} =\,-{\rm u}_1^-\;.
\ee

\subsubsection{Weak coupling limit of the GRWA}

To first order in $\tilde\alpha$,  $\mathsf{L}^{1}_n(\tilde\alpha)\simeq n+1$ and $\mathsf{L}^0_n(\tilde\alpha)\simeq 1$. It follows that to first order in $g$
\be{}
\tilde\Delta_{nn}&\simeq \Delta\;,\\
\omega_{q,n}&\simeq\omega_q\;,\\
c^\pm_n&\simeq \sqrt{\frac{\omega_q\pm\epsilon}{2\omega_q}}\;,\\
\Omega_{n}&\simeq \tilde\Delta_{nn-1}\simeq \Delta\sqrt{n\tilde\alpha}\;,\\
\delta_n&\simeq\omega_q-\omega_r\;.
\ee
As a result, to first order in $g$, the spectrum acquires a form similar to that of the RWA, Eq.~\eqref{JCMspectrum}. 
\begin{equation}
\begin{aligned}
\omega_0 &\simeq\;-\omega_q/2\;,\\
\omega_{2n-1}&\simeq\; \left(n-\frac{1}{2}\right)\omega_r 
-\frac{1}{2}\sqrt{\left(
\omega_q
-\omega_r\right)^2+n(\Delta\tilde\alpha)^2}\;,\\
\omega_{2n}&\simeq\; 
\left(n-\frac{1}{2}\right)\omega_r 
+\frac{1}{2}\sqrt{\left(
\omega_q
-\omega_r\right)^2+n(\Delta\tilde\alpha)^2}
\;,
\end{aligned}
\end{equation}
the difference being that the term $\sqrt{\tilde\alpha}\Delta=2\Delta g/\omega_r$ is replaced in the RWA by $2g_x=2\Delta g/\omega_q$.
The gap of the effective qubit reads
\be{}
\omega_{10} =\frac{\omega_q}{2}+\frac{\omega_r}{2}-\frac{1}{2}\sqrt{\left(
\omega_q
-\omega_r\right)^2+(\Delta\tilde\alpha)^2}\;.
\ee
The weak-coupling limit of the matrix elements for generic bias $\epsilon$ are given by Eq.~\eqref{matrix_elements_GRWA_app} and read
\be{}
Q_{L,01} &=\, \frac{2 g}{\omega_r}{\rm u}_1^-\frac{\Delta}{\omega_q}+ {\rm v}_1^-\;,\\
Q_{R,01} &=\,-{\rm u}_1^-\frac{\Delta}{\omega_q}\;,
\ee
where
\be{u_v_GRWA_weak_coupling}
{\rm u}_1^- &=\;\frac{\omega_q-\omega_r - \sqrt{(\omega_q-\omega_r)^2+(\Delta\tilde\alpha)^2}}{\sqrt{\left(
\omega_q-\omega_r - \sqrt{(\omega_q-\omega_r)^2+(\Delta\tilde\alpha)^2}
\right)^2+(\Delta\tilde\alpha)^2}}\\
{\rm v}_1^- &=\;\frac{-\Delta\tilde\alpha}{\sqrt{\left(
\omega_q-\omega_r - \sqrt{(\omega_q-\omega_r)^2+(\Delta\tilde\alpha)^2}
\right)^2+(\Delta\tilde\alpha)^2}}\;.
\ee
\end{widetext}


%

\end{document}